\documentclass[preprint]{aastex}

\usepackage{amsmath}
\usepackage{amssymb}
\usepackage{latexsym}

\newcommand{\kms}{km\,s$^{-1}$}

\begin{document}
\title{Observations of ubiquitous compressive waves in the Sun’s chromosphere}

\author{Richard J. Morton\altaffilmark{1}, Gary Verth\altaffilmark{1}} 
\affil{Solar Physics and Space Plasma Research Centre
(SP$^2$RC), The University of Sheffield, Hicks Building, Hounsfield
Road, Sheffield S3 7RH, UK}

\author{David B. Jess\altaffilmark{2}, David Kuridze\altaffilmark{2}}

\affil{Astrophysics Research Centre, School of Mathematics and Physics, Queen’s University, Belfast, BT7 1NN, UK.}

\author{Michael S. Ruderman\altaffilmark{1}, Mihalis Mathioudakis\altaffilmark{2}, Robertus Er\'elyi*\altaffilmark{1}}

\date{Received /Accepted}
\begin{abstract}
The details of the mechanism(s) responsible for the observed heating and dynamics of the solar 
atmosphere still remain a mystery. Magnetohydrodynamic (MHD) waves are thought to play a vital role in 
this process. Although it has been shown that incompressible waves are ubiquitous in off-limb solar 
atmospheric observations their energy cannot be readily dissipated. We provide here, for the first time, 
on-disk observation and identification of concurrent MHD wave modes, both compressible and 
incompressible, in the solar chromosphere. The observed ubiquity and estimated energy flux associated 
with the detected MHD waves suggest the chromosphere is a vast reservoir of wave energy with the 
potential to meet chromospheric and coronal heating requirements. We are also able to propose an upper 
bound on the flux of the observed wave energy that is able to reach the corona based on observational 
constraints, which has important implications for the suggested mechanism(s) for quiescent coronal 
heating. \\

\vspace{0.2cm}
*To whom correspondence should be addressed. Email: robertus@sheffield.ac.uk
\end{abstract}


\section{Introduction}
The quiet chromosphere is a highly dynamic region of the Sun’s atmosphere. It consists of many small-
scale, short-lived, magnetic flux tubes (MFTs) composed of relatively cool ($10^4$ K) plasma [1]. The 
quiet chromosphere may be divided into two magnetically distinct regions [2]. One is the network, which 
generally consists of open magnetic fields outlined by plasma jets, i.e., spicules and mottles, which 
protrude into the hot ($10^6$ K) corona. The other region is the inter-network that is populated by 
fibrils that outline magnetic fields, joining regions of opposite polarity hence are closed within the 
chromosphere [1,3,4]. The possible importance of chromospheric dynamics on the upper atmosphere has 
recently been highlighted by joint Hinode and Solar Dynamic Observatory (SDO) [5] and ground-based 
observations by the Swedish Solar Telescope [6] showing a direct correspondence between 
chromospheric structures and plasma that is heated to coronal temperatures.

Magnetohydrodynamic (MHD) waves have long been a suggested mechanism for distributing magneto-
convective energy, generated below the solar surface, to the upper layers of the Sun's atmosphere [7,8]. 
Incompressible MHD wave modes, characterised by the transverse motions of the magnetic field lines, 
have been observed to be ubiquitous throughout the solar chromosphere [9-11] and corona [12-14]. 
However, incompressible wave energy is notoriously difficult to dissipate under solar atmospheric 
conditions [5] requiring very large Alf\'en speed gradients occurring over short (sub-resolution) spatial 
scales to achieve the necessary rate of plasma heating.

In contrast, compressible MHD waves are readily dissipated by, e.g., compressive viscosity and thermal 
conduction [15]. Active region fast compressible MHD waves have recently been identified at both 
photospheric and coronal heights [16-18]. There has been some tentative evidence for their existence in 
the quiet chromosphere [19] but now we establish their ubiquity in this paper, making them a possible 
candidate for plasma heating in the quiescent corona. 

  We demonstrate here the presence of concurrent, ubiquitous, fast compressive MHD waves and 
incompressible MHD waves in the on-disk solar chromosphere. Measurements of the wave properties 
(i.e., period, amplitude, phase speed) for both compressible and incompressible modes are provided. 
Further, we derive estimates for the flux of wave energy by combining advanced MHD wave theory and 
the observed wave properties. These estimates hint that the observed waves could play a crucial role in 
the transport of magneto-convective energy in the solar atmosphere.  
  
\section{Results}

\begin{figure}[!htp]
\centering
\includegraphics[scale=1.]{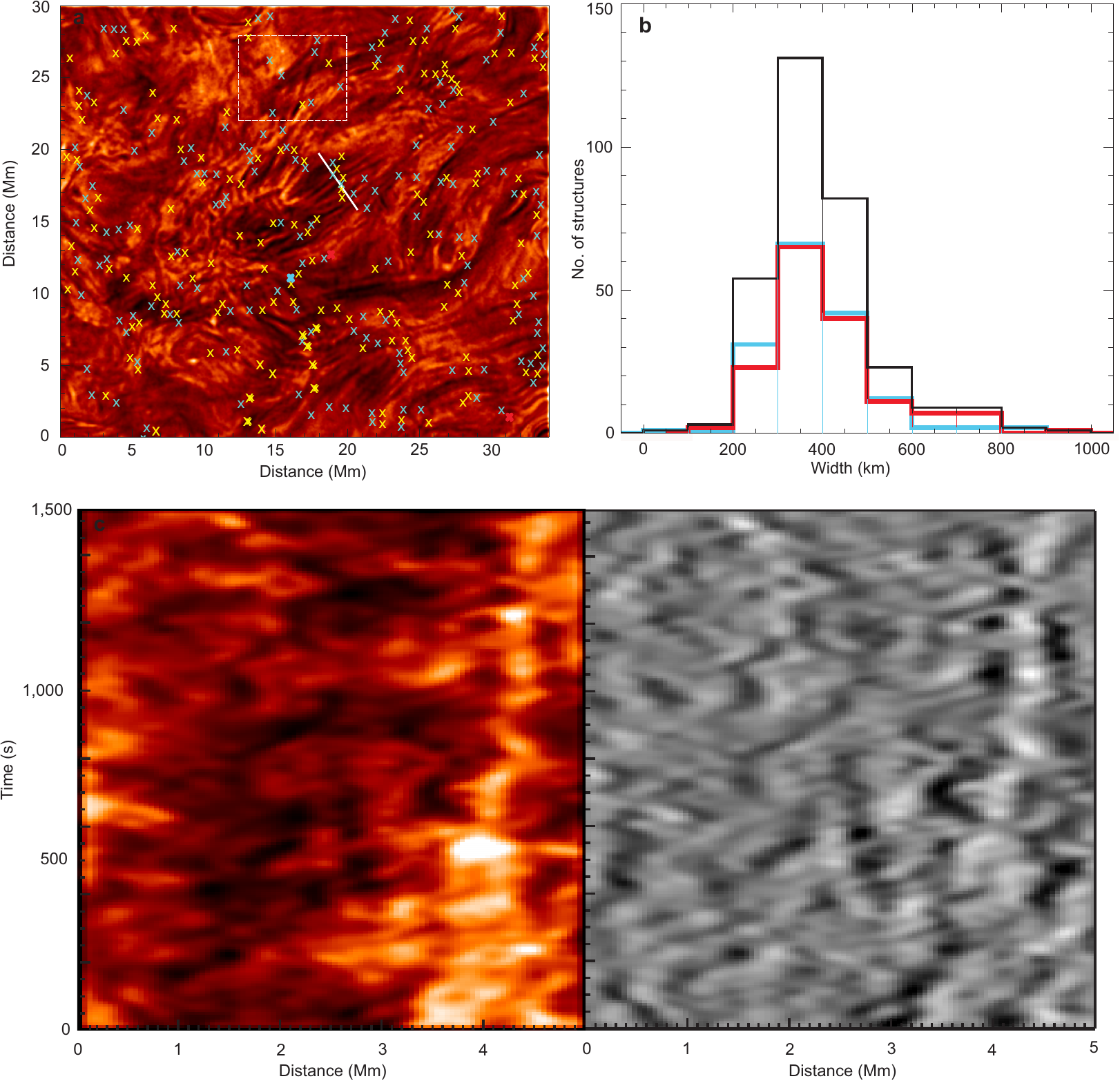}
\caption{ Observed chromospheric structures and their transverse motions. (a) The 
ROSA Hα field of view at t=0~s. Blue and yellow crosses mark positions of identified dark and white 
chromospheric MFTs, respectively. The boxed region is where the chromospheric structures in Fig.~4 are 
observed. (b) Histogram of measured MFT widths. Blue and red lines correspond to the dark and bright 
MFTs, respectively. The black line is the combined result with a mean width of $360\pm120$~km. (c) 
Time-distance plot (left) and median filtered version (right) taken across a group of chromospheric 
structures. The cross-cut used is shown in (a). The given distance starts from the bottom of the cross-
cut and the time is in seconds from the beginning of the data set. The dark and white linear tracks are 
the transverse motions of the structures. }
\end{figure}

\subsection{The chromospheric fine structure}
High spatial (150~km) and temporal resolution (8~s) observations with the Rapid Oscillations in the Solar 
Atmosphere (ROSA) imager [20] have allowed us to resolve the chromospheric structures and measure 
fine-scale wave-like motions therein. The data are obtained close to disk centre with a narrowband 
0.25~${\AA}$ H$\alpha$ core (6562.8~${\AA}$) filter on the 29 September 2010 (see Methods for 
further information on the processing of the observational data). The field of view covers a 
34~Mm$\times$34~Mm 
region of the typical quiet Sun (Fig. 1, Supplementary Fig. S1). The H$\alpha$ core images are known to 
show the mid- to upper chromosphere where the magnetic pressure dominates the gas pressure [3,21]. 
The images (Supplementary Movie S1) show that the on-disk chromosphere is dominated by hundreds of 
bright and dark fine structures, all displaying highly dynamic behaviour. The elongated structures are the 
fibrils and shorter structures are the mottles. Recent advances in the modelling of H$\alpha$ line 
formation suggest the dark structures are regions of enhanced density that closely follow the magnetic 
field structure [22]. The nature of the bright fine structures are less clear, although they are suspected to 
be similar to the dark features except for differences in their gas pressure [23]. In Fig.1 we show the 
measured widths of over 300 bright and dark structures, and the results demonstrate the geometric 
similarity between these structures. 

The observed structures can be considered as discrete over-dense MFTs outlining the magnetic field, 
embedded in a less dense ambient plasma. They are short lived (1-5 minutes) but fortunately survive 
long enough to allow measurements of important wave parameters, e.g., velocity amplitude and 
propagation speed. The observations show that similar structures re-occur in the same regions either 
continuously or even tens of minutes after previous structures have faded from the field of view 
(Supplementary Movies S1 and S2), suggesting that the lifetime of the background magnetic field is much 
longer than the plasma flows that generate the visible structuring. 

Waves may be present in the atmosphere even when the magnetic field is not outlined by an intensity 
enhancement; however, they are far more difficult to identify directly. These intensity enhancements 
suggest a greater plasma density along the waveguide relative to the surrounding atmosphere, which has 
important physical effects on the properties of the wave. Mathematically, it is well known that plasma 
structuring in solar waveguides, even if modelled as relatively simple over-dense cylindrical MFTs, can 
produce an infinite set of orthogonal eigenmodes [24,25]. This is in stark contrast to only three possible 
modes of a homogeneous magnetized plasma (Alfv\'en, fast and slow magnetosonic modes). Although the 
full MHD spectrum of permitted wave modes in realistic inhomogeneous solar MFTs is rich and complex, 
in practice, due to limits in spatial resolution, most of the wave power observed so far has been confined 
to the low order azimuthal wave number modes, i.e., torsional Alfv\'en, sausage and kink. Higher order 
fluting modes predicted by theory, to our knowledge, have yet to be detected in solar data. It has, 
however, been suggested by numerical simulations that different MHD wave modes should occur 
simultaneously in chromospheric MFTs [26]. In this current work, we identify such concurrent wave 
modes.

\begin{figure}[!h]
\centering
\includegraphics[scale=1.]{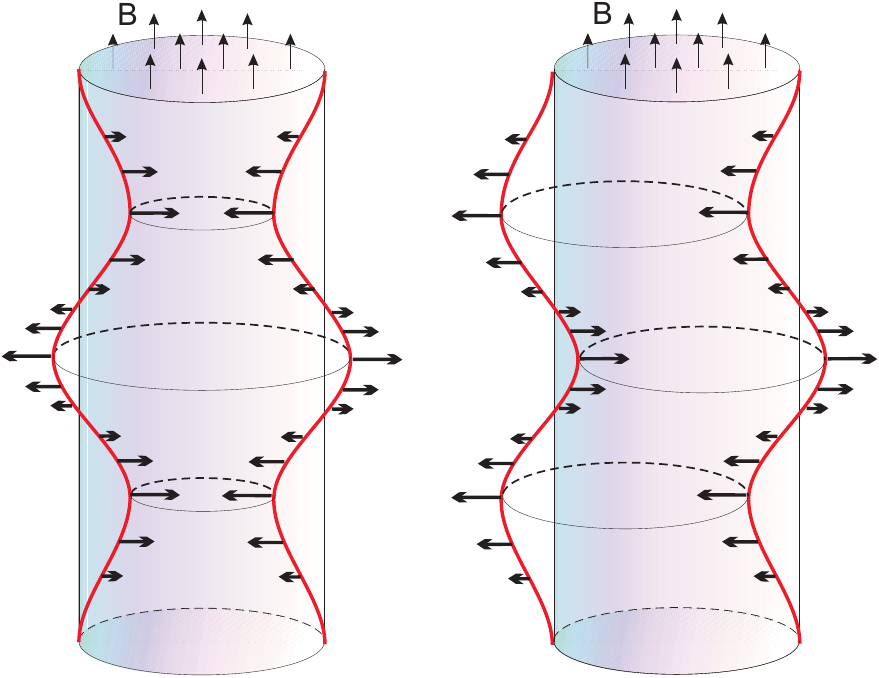}
\caption{ Schematic diagrams of MHD waves in a magnetic flux tube. The MHD wave 
modes supported by the chromospheric fine structure can be modelled as a cylindrical magnetic 
waveguide. The fast MHD sausage wave (left) is characterised by a periodic contraction and expansion of 
the MFTs cross-section symmetric about the central axis, stretching and squeezing the magnetic field. 
This, in turn, induces a decrease and increase in the plasma pressure inside the MFT that changes the 
intensity of the observed MFT in the Hα bandpass. The fast MHD kink wave (right) displaces the central 
axis of the MFT. The red lines show the perturbed waveguide and thick arrows show the velocity 
amplitudes. The thin arrows labelled B show the direction of the background magnetic field. }
\end{figure}

\subsection{Incompressible transverse waves}
To analyse the waves, a cross-cut is placed perpendicular to a chromospheric structure’s axis and time-
distance plots are constructed (e.g., Fig. 1, Supplementary Figs. S2, S3). We find that the chromospheric 
structures support two concurrent orthogonal MHD wave modes. The first type is the transverse motion 
of the structure’s axis (Fig. 2), which is ubiquitous and even visible by eye (Supplementary Movies S1-S4). 
In terms of the over-dense magnetic waveguide model we interpret these motions as the highly 
incompressible MHD fast kink wave [24,25]. The transverse displacements appear predominantly as 
linear, diagonal tracks in the time-distance plots, although occasionally leaving sinusoidal tracks (see 
Methods for further details). The properties of over a hundred representative transverse motions were 
measured and they reveal that the typical displacement amplitudes are 315$\pm$130~km with velocity 
amplitudes of 5-15~\kms (Fig. 3). For isolated chromospheric structures (e.g., those in Fig. 4 and 
Supplementary Fig. S2) a Gaussian can be reliably fitted to the cross-sectional flux profile to obtain the 
transverse displacement and velocity amplitudes, and the period (see Methods for further details). A 
linear fit to the diagonal tracks is undertaken when a Gaussian fit cannot be performed. The number of 
structures to which a Gaussian fit can be applied is 47; the remainder of the fits are linear. Using cross-
correlation, the propagation speeds can also be measured and values in the range 40-130~\kms are 
obtained (Supplementary Table S1). We note that the amplitude and period of the waves observed here 
are similar to those reported in limb observations of spicules [10]. The amplitudes of the transverse 
motions are of the order of, or greater than, the waveguide width (360~km), implying that they are fast 
kink wave modes in the non-linear regime.

\begin{figure}[!h]
\centering
\includegraphics[scale=1.]{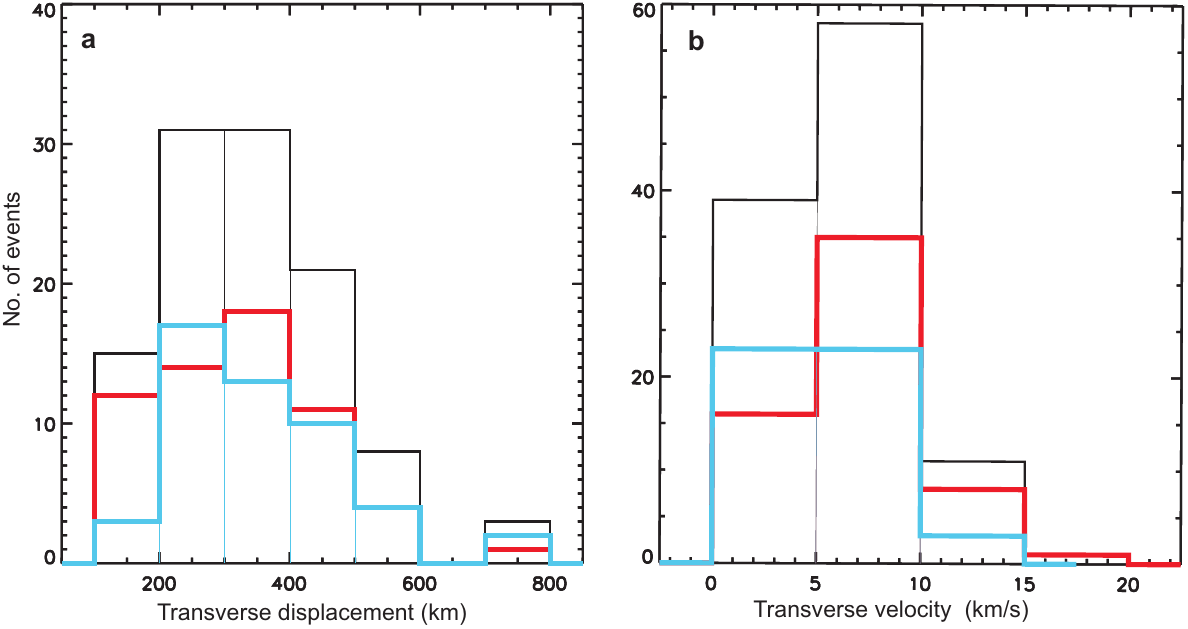}
\caption{Measured properties of transverse displacements and velocity amplitudes for 
the fast MHD kink mode. (a) The distribution of transverse displacements of 103 chromospheric 
structures measured using linear fits to the observed tracks. The blue and red lines correspond to dark 
and bright structures, respectively. The black line shows the combined results with a mean displacement 
of $315$~km and standard deviation of $130$~km. (b) Velocity amplitudes for both bright and dark 
chromospheric structures. The transverse velocity has a mean value of $6.4$~\kms with a standard 
deviation of $2.8$~\kms. The dark and bright structures show a similar distribution. }
\end{figure}

\subsection{Fast compressible waves}
The second type of dynamic behaviour in chromospheric structures (Fig. 2), concurrent with the 
transverse displacement (Figs. 4, 5, Supplementary Figs. S2, S4), is a periodic increase and decrease of 
the structures’ intensity. In conjunction, there is an anti-correlated contraction and expansion of the 
structures’ visible cross-section. Both the intensity and width perturbations are obtained from the 
Gaussian fits. This perturbation is found to propagate along chromospheric structures with speeds much 
greater than the sound speed and is comparable to the local Alfv\'en speed (see Methods for further 
details). We interpret the observed periodic changes in intensity from MHD wave theory [24,25] as being 
consistent with the compressive properties of the fast MHD sausage mode. The visible lifetime of the 
chromospheric structures are also comparable to the fast sausage and kink modes’ period, which means 
that observing multiple wave cycles is rather difficult. However, the fast propagating intensity 
disturbances can also be identified in time-distance diagrams that are constructed when a cross-cut is 
placed parallel to a structure’s axis (see, Fig.~5, Supplementary Figs.~S4, S5, S6). These events are seen 
to occur continuously all over the field of view (Supplementary Movies S3 and S4). A table of measured 
phase speeds and periods are provided in Supplementary Table S2.

Interestingly, the sausage mode has distinct regimes that define whether the majority of wave energy can 
be trapped in the vicinity of a density enhancement along the magnetic field or it leaks away [24,25,27]. 
To investigate whether the observed fast MHD sausage modes are mostly trapped or leaky we have to 
estimate if the product of wavenumber (k) and waveguide half-width (a) is a sufficiently small parameter.  
For the observed chromospheric structures the sausage mode would have to have $ka>0.2$ to be 
trapped but we observe, e.g., for the example in Fig.~4, that $ka\sim0.08\pm0.03$ (see Supplementary 
Information for further details). Therefore, the observed sausage modes appear to be in the leaky regime, 
further enhancing their already dissipative nature.  It is important to note that leaky sausage modes, with 
no damping due to physical effects such as compressive viscosity or radiation, would still radiate energy 
away from the waveguide with decay times on the order of the period.  

Now, let us estimate the energy flux for the observed MHD waves using the waveguide model. The 
polarisation relations for a particular wave mode allow us to determine the perturbations of all the 
relevant physical quantities, if we know the perturbation of one [24,25]. This way we obtain an estimate 
for the amplitude of the perturbations we cannot measure directly otherwise, e.g. the magnetic field 
fluctuations (see Supplementary Information for further details). The polarisation relations need to be 
supplemented with typical values of chromospheric plasma parameters (e.g., density [1,22], 
$\rho=3\times10^{-10}$~kg/m$^3$, temperature [22,23], $T=10,000$~K and magnetic field strength 
[28], $B=10-30$~Tesla) and the amplitudes and phase speeds measured here. The time-averaged wave 
energy flux in the chromospheric 
plasma is composed of two dominant components, the kinetic and the magnetic energy flux, 
$<E>=c_{ph}[\rho v^2+b^2/μ_0]/4$, where $v$ and $b$ are the amplitudes of the velocity and 
magnetic perturbations, respectively, $c_ph$ is the phase speed and $μ_0$ is the magnetic permeability 
of free space. The thermal 
energy contributions from each mode are found to be negligible due to the magnetic pressure being at 
least an order of magnitude greater than the gas pressure in the chromosphere. The estimated energy 
flux of the incompressible fast kink mode is $<E_k>=4300\pm2700$~W/m$^2$, which is comparable to 
that estimated in off-limb spicules [9,10]. The estimated wave energy flux for the compressible fast MHD 
sausage mode is $<E_s>=11700\pm3800$~W/m$^2$ per chromospheric structure for apparent radial 
velocity amplitudes of $v~1-2$~\kms. The details of the wave energy calculation applicable to the 
observed inhomogeneous structured plasma are given in the Supplementary Methods, Supplementary 
Figs. S7, S8 and Supplementary Table S3.

\begin{figure}[!h]
\centering
\includegraphics[scale=1.]{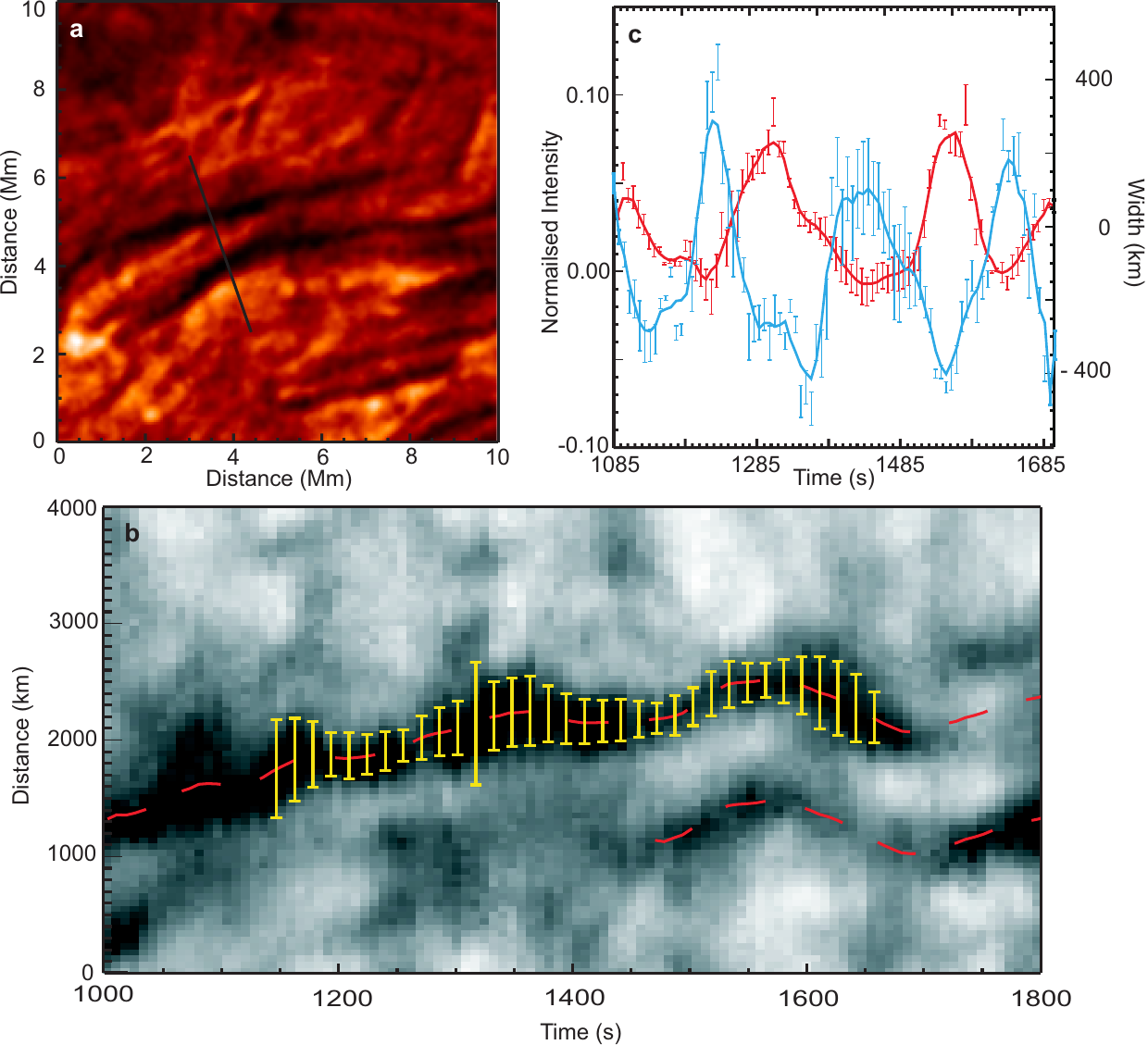}
\caption{Concurrent wave modes in chromospheric structures. (a) Typical ROSA 
H$\alpha$ example of a pair of relatively large dark flux tubes at $t=1536$~s measured from the 
beginning of the data series. (b) Time-
distance plot revealing the dynamic motion. The position of the cross-cut is shown in (a), with the given 
distance starting at the top of the cross-cut. Times are given in seconds from the start of the data set. 
The results from a Gaussian fitting are over-plotted and show the non-linear fast MHD kink wave (red 
line shows the central axis of the structure) and the fast MHD sausage mode (yellow bars show the 
measured width of structure). The transverse motion has a period of $232\pm8$~s and we detect multi-
directional propagating transversal wave trains in the MFT travelling with speeds of $71\pm22$~\kms 
upwards and $87\pm26$~\kms downwards (for further details see, Supplementary Fig. S6). The typical 
velocity amplitudes are 5~\kms. The fast MHD sausage mode has a period of $197\pm8$~s, a phase 
speed of $67\pm15$~\kms and apparent velocity amplitudes of $1-2$~\kms (c) Comparison of MFTs 
intensity (blue) and width (red) perturbations from the Gaussian fitting. The data points have been fitted 
with a smoothed 3-point box-car function. The observed out-of-phase behaviour is typical of fast MHD 
sausage waves. The error bars plotted are the one-sigma errors on each value calculated from the 
Gaussian fitting.}
\end{figure}

A further comment on the wave energy flux is required. The over-dense cylindrical waveguide supports 
different MHD wave modes, where some of the energy of trapped modes is concentrated near the 
waveguide. Consequently, the consideration of collections of over-dense waveguides occupying the 
chromosphere results in a much more complex calculation of the total energy flux. The derived value 
does not take into account the following: The wave energy of leaky modes is not confined within the flux 
tube, with a comparable amount of perturbation energy outside the tube. This means calculation of the 
internal wave energy flux only, leads to an underestimate; Only a linear approach for the wave energy 
flux is given. The transverse motions are generally observed to be strongly non-linear, so we are 
underestimating the wave energy flux by ignoring these non-linear terms; The true nature of the 
chromospheric fine structure is difficult to determine at present. The homogeneous cylindrical model is a 
simplifying assumption and does not account for the true geometry of the structure and/or any radial 
structuring of the plasma. All the above factors could influence the energy estimates. Calculating the 
total wave energy in a plasma such as the chromosphere is complex and will require further observations 
and full 3-dimensional MHD numerical simulations before improved values for wave energy flux can be 
obtained.

\begin{figure}[!h]
\centering
\includegraphics[scale=1.]{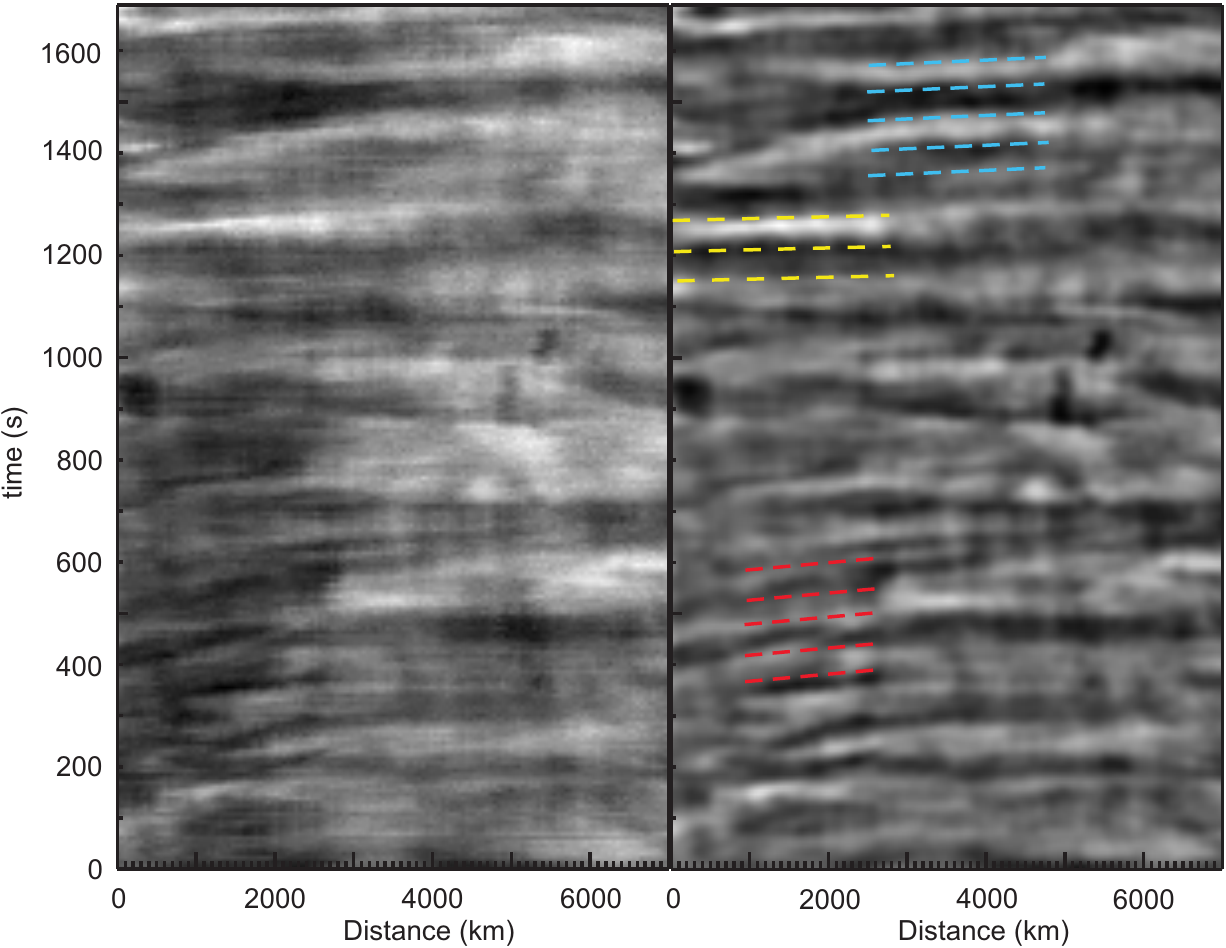}
\caption{Fast propagating intensity perturbations (a) A typical time-distance diagram 
obtained by placing a cross-cut parallel to a chromospheric structure. This particular cross-cut is 
perpendicular to the cross-cut shown in Fig.~1a. (b) The same time-distance plot is shown as in (a) but it 
has been subject to a high-pass Atrous filter. The result reveals periodic variations in intensity that 
propagate along the cross-cut, hence the chromospheric structure. Measuring the gradients of the tracks 
created by the intensity perturbations across the time-distance plots provides propagation velocities. The 
coloured, dashed lines highlight three examples of these periodic variations. The measured propagation 
velocities are $78\pm9$~\kms (red), $270\pm135$~\kms (yellow) and $135\pm17$~\kms (blue). The 
amplitude of the perturbation in intensity is $\leq 10\%$ of the background value, similar to that found 
from the Gaussian fitting (Fig. 4). }
\end{figure}

\section{Discussion}

Knowing the eventual fate of the wave energy observed in the chromosphere is vital in assessing which (if 
any) wave-based heating models are likely candidates for explaining the long-standing solar atmospheric 
heating puzzle [29-35]. Using the measured properties of the chromospheric structures, we estimate (see 
Supplementary Information for further details) that a maximum of $4-5\%$ of the chromospheric volume 
is connected by MFTs that protrude into the corona. This estimate of connectivity between the 
chromosphere and corona is subject to observational constraints but the given value may not change 
significantly with improved spatial resolution. 

The estimated connectivity suggests that the total energy flux across the solar surface able to reach the 
corona from the chromosphere is $\sim170\pm110$~W/m$^{-2}$ for the incompressible transverse 
motions and $\sim460\pm150$~W/m$^2$ for compressive motions. The estimate for chromospheric 
transverse wave energy able to reach the lower corona agrees with the energy estimates in the same 
period range (100-500~s) from off-limb SDO quiet Sun observations [14], which could suggest that these 
waves remain relatively undamped on their journey. This can be explained by the known difficulty of 
dissipating these incompressible waves [15]. Since it is expected that the majority of field lines observed 
here remain closed (i.e., the fibrilar structure), this suggests that the observed chromospheric wave 
energy remains largely in the chromospheric volume to be dissipated in a different manner [33-35]. On 
the other hand, the eventual fate of the chromospheric fast compressive waves is still unclear. However, 
our results imply that the waves could play a significant role in meeting chromospheric 
($\sim10$~kW/m$^2$ [36]) and quiet coronal heating requirements ($100-200$~W/m$^2$ [36,37]). 
Further work is needed to: 
provide more representative statistics on typical perturbation amplitudes of the fast compressive waves, 
which will lead to improved estimates for the energy flux of the waves; search for coronal counterparts of 
the sausage MHD waves in the corona; and assess how the leaky nature of the waves affects the wave 
energy transmission to and dissipation in the lower corona. 

The observations presented here demonstrate that the on-disk quiet chromosphere, which is dominated 
by fibrilar structure, supports ubiquitous incompressible transverse waves. The properties of the 
transverse waves observed agree with those reported to exist in limb spicules [10], which are magnetic 
waveguides that penetrate into the low corona. This points towards a close connection between the disk 
and limb structures even if their magnetic topology (i.e., open or closed magnetic fields) is potentially 
different. In addition, the observation and identification of ubiquitous, fast propagating, compressible 
MHD waves demonstrate that the chromosphere is replete with MHD wave energy. The ubiquity of both 
the incompressible and compressible waves, in conjunction with our initial estimates of wave energy, 
gives further credence to models of a predominantly wave-based quiescent atmospheric heating 
mechanism. 
  
\section{Methods}

\subsection{Observational set-up and data processing}
The data was obtained at 15:41-16:51~UT on 29 September 2010 with the Dunn Solar Telescope at 
Sacremento Peak, USA. The six-camera system Rapid Oscillations in the Solar Atmosphere (ROSA) was 
employed. ROSA observed a 69′′.3 by 69′′.3 region of the quiet solar atmosphere in Hydrogen $\alpha$ 
(6562.7~${\AA}$). The H$\alpha$ filter is centred at the line core and a narrow bandpass (0.25~${\AA}$) 
is employed to ensure that photospheric contributions from the line wings are minimised (for details on 
H$\alpha$ line formation see, e.g., [refs. 3, 21, 22]). During the observations, high-order adaptive 
optics [38] were used to correct for wave front deformation in real time. The seeing conditions were 
good. The data suffers from a period of poor seeing during the middle of the run lasting 300~s. The 
frames in this period of bad seeing are not used for analysis. The H$\alpha$ data was sampled at 2.075 
frames s$^{-1}$ and image quality improved through speckle 
reconstruction [39] utilising a 16−1 ratio, giving a final cadence of 7.68~s. To ensure accurate co-
alignment between frames, the narrowband time series were Fourier co-registered and de-stretched 
using vectors derived from longer-lived broadband structures [40].

\subsection{Gaussian fitting of chromospheric structures}
We discuss here the procedure used to observe the periodic phenomena in the chromospheric magnetic 
structures. First, we place cross-cuts perpendicular to the structure’s central axis and obtain time-
distance diagrams as shown in Figs.~1, 4 and Supplementary Fig. S2. In each time slice, t, of the time-
distance plot we fit a Gaussian profile to the cross-sectional flux profile, $F(x,t)$, of the chromospheric 
magnetic structure, where the Gaussian fit is given by
\begin{equation}
F_{fit}(x,t)=f_0 (t)\exp\left(-\frac{(x-f_1 (t))^2}{2f_2^2 (t)}\right)+f_3 (t)
\end{equation}
Here, $f_0(t)$ is the peak flux, $f_1(t)$ is the central position of the Gaussian, $f_2(t)$ is the Gaussian 
width and $f_3(t)$ is the background flux. The obtained fit parameters relate to the periodic motions: 
$f_0(t)$ gives intensity (or flux) perturbation; f1(t) gives the displacement of the structure’s central axis, 
i.e., the transverse waves; $f_2(t)$ gives the full-width half-maximum (FWHM), $FWHM=2f_2\sqrt{2\ln2}$, 
of the structure’s visible cross-section. Before measuring the properties (e.g., period, velocity amplitude, 
etc.) of each series $f_0$, $f_1$..., each time series is de-trend by fitting a cubic function and 
subtracting the trend. The periods of the wave motion are obtained by using wavelet analysis. Examples 
of structures demonstrating periodic phenomenon derived from the Gaussian fits are shown in Fig.~4 and 
Supplementary Fig.~S1. The Gaussian fit is applied only to isolated chromospheric flux tubes to avoid the 
influence of the chromospheric structures crossing each other, which would lead to changes in both 
intensity and width. All errors provided with respect to the results obtained from the Gaussian fitting (i.e., 
those in Fig.~4, Supplementary Table S1) are one standard deviation.

\subsection{Measuring transverse amplitudes}
Here we discuss how the transverse amplitudes are obtained for the histograms in Fig.~3. In the time-
distance diagrams a large number of diagonal tracks (Fig.~1, Supplementary Fig. S3), with varying 
gradients, and a smaller number of sinusoidal tracks (e.g., Fig~4, Supplementary Fig. S2) can be 
identified. These are the transverse motions of the chromospheric structures. The transverse amplitudes 
are obtained by measuring the length of the bright or dark diagonal tracks and the time averaged velocity 
amplitudes are the gradients of the tracks. The Gaussian fit is not used here for determining the centre of 
the chromospheric structure. The spatial position of the centre of the chromospheric structures can be 
measured to the nearest pixel by locating the maximum/minimum value of intensity, which provides an 
error of 50~km on position. The time coordinate for each position of the chromospheric structure can 
only be known to within a range of 7.68~s, hence the error is $dt\pm3.84$~s. For the measured values 
of displacement amplitude given in Fig.~3 the median value of the error is $\sim13\%$ of the given 
value. The velocity amplitude has a median error of $22~\%$ of the given value, which equates to 
$1.3$~\kms and a standard deviation on the error of $1.1$~\kms. 

   Calculating the velocity amplitude from the straight line fits only provides the time-averaged value of 
the amplitude. If a number of oscillatory cycles can be observed, then the following relation is used to 
find the maximum velocity amplitude,
\begin{equation}
v_r=\frac{\partial\xi_r}{\partial t}=\frac{2\pi}{P \xi_r}
\end{equation}
Here $\xi_r$ is the transverse displacement (given by $f_1$ in the Gaussian fit), $v_r$ is the radial 
velocity perturbation and $P$ is the period of the oscillation. Only a small percentage of the observed 
oscillations show full cycles, i.e., sinusoidal motions in the time-distance plots. The reason why we do 
not see full cycles is due to the relatively short lifetimes (1-5 minutes) of the chromospheric magnetic 
structures compared to a typical period of the fast kink waves (3 minutes). A similar behaviour is 
observed for limb spicules [9,10].

\subsection{Amplitudes of the fast MHD sausage mode} 
Here we describe how we estimate the velocity component of the fast MHD sausage mode from the 
changes in the FWHM as established by the Gaussian fitting outlined in the Methods section in the main 
paper. We assume that the compressive motion is a symmetric perturbation about the central axis of the 
waveguide. The FWHM is a measure of the chromospheric structure’s visible diameter. We make the 
assumption that the displacement in the radial direction occurs at the same rate as the change in the 
value of the FWHM. The radial velocity amplitude is given by Equation 2 and we assume $\xi_r \propto 
FWHM/2$. Equation~2 can only be used when a number of cycles of the fast sausage mode are observed. 
The value for $v_r$ is the value of maximum amplitude. The time-averaged value is given by $<~vr~>= 
v_r/\sqrt{2}$. We note that the assumption regarding the relationship between FWHM and the radial 
velocity may not hold true if the change in the visible cross-section does not accurately represent the 
physical change in cross-section. Detailed simulations of waves in plasmas and their influence on 
H$\alpha$ line formation may be needed to resolve such issues.

\subsection{Fast propagating intensity disturbances}
The spatial resolution of ROSA ($\sim150$~km) is close to the widths of the chromospheric flux tubes 
($\sim350$~km), making the joint detection of periodic changes in intensity and the width of flux tubes 
difficult to observe. However, it is possible to identify the fast compressive wave mode through 
identification of periodic, propagating intensity disturbances (Fig.~5, Supplementary Fig. S4, S6). The 
intensity variations seen in the filtered data is $\leq10\%$ of the median intensity of the unfiltered 
intensity. This variation in intensity is in agreement with the values obtained from the Gaussian fits 
(compare Fig.~4 and Supplementary Fig.~S2). From these filtered time-distance plots we are able to 
measure propagation speeds and periods of the observed periodic disturbances (see Supplementary 
Table~S2). We identify 10 regions in which the properties of the fast propagating features can be 
measured (Supplementary Fig.~S6).

    We are confident the observed perturbations are due to the fast sausage modes as the periodic 
features have a large spatial coherence ($>1000$~km - see Supplementary Movies S3 and S4). This 
behaviour is somewhat expected due to the leaky nature of the oscillations (see Supplementary 
Information for further details). In contrast, the observed kink motions have a small spatial coherence 
(see, Fig.~1), with neighbouring flux tubes showing little or no coherence.

\subsection{Measuring phase speeds} 
To measure the phase speed for the observed waves in chromospheric structures, we have to be able to 
track the periodic variation in, i.e., displacement, intensity, at different positions along the structure. We 
place cross-cuts at different positions along the structure’s central axis and fit a Gaussian to the 
observed sinusoidal variations. The time lag between the perturbed quantities obtained from the 
Gaussian fit in the different cross-cuts is obtained by using cross-correlation to calculate the shift 
between two signals. The phase speed is given as the average velocity determined from the time lags 
between different cross-cuts (see Supplementary Information for further details) and the error is the 
standard deviation.

 The propagating nature of the waves is also seen by plotting the transverse displacement for different 
cross-cuts (as measured with the Gaussian fits). An example of this is shown in Supplementary Fig. S5, 
where the propagation of the transverse wave shown in Fig. 4 can be seen. The distance between the 
displayed cross-cuts is $\sim150$~km. We note that it is possible that the phase speed varies as a 
function of distance along the structure due to density and magnetic field stratification; however, a 
combination of the technique used and the spatial and temporal resolution of the current observations 
does not allow us to determine this variation of phase speed. 

  The phase speeds for the fast compressive disturbances are calculated differently. Once a periodic 
propagating disturbance is identified in time-distance plot (e.g., Fig.~5, Supplementary Fig.~4), the 
minimum or maximum valued pixel in time (depending on whether the maximum or minimum of the 
intensity perturbation is being measured) in each spatial slice is located. The gradient of a linear fit to the 
minimum/maximum values gives the phase speed. The errors on the spatial locations are $50$~km and 
the time coordinates are $3.84$~s. As the propagating disturbances are seen over relatively long 
distances for short periods of time, the largest source of error comes from the uncertainty in time.

\section{References and Notes: }
\noindent1. Beckers, J. Solar spicules. Sol. Phys. 2, 367-433 (1968)\\
2. Dowdy, J., Rabin, D., Moore, R. On the magnetic structure of the quiet transition region. Sol. Phys. 105, 
35-45 (1986)\\
3. Rutten, R., The quiet-Sun photosphere and chromosphere, Phil. Trans. Roy. Soc A, 370, 3129-3150 
(2012)\\
4. Schrijver, C. et al. Large-scale coronal heating by the small-scale magnetic field of the Sun. Nature 
394, 152-154 (1998)\\ 
5. De Pontieu, B. et al. The origins of hot plasma in the solar corona. Science 331, 55-58 (2011)\\
6. Wedemeyer-Böhm, S. et al. Magnetic tornadoes as energy channels into the solar corona. Nature, 486, 
505-508 (2012)\\
7. Schatzman, E. The heating of the solar corona and chromosphere. Annal. d'Astrphys. 12, 203-218 
(1949)\\
8. Osterbrock, D. The heating of the solar chromosphere, plages and corona by magnetohydrodynamic 
waves. Astrophys. J. 134, 347-388 (1961)\\
9. De Pontieu, B. et al. Chromospheric Alfv\'enic waves strong enough to power the solar wind. Science 
318, 1574-1577 (2007)\\
10. Okamoto, T., De Pontieu, B. Propagating waves along spicules. Astrophys. J. 736, L24 (2011)\\
11. Jess, D. et al. Alfv\'en waves in the lower solar atmosphere. Science 323, 1582-1585 (2009) \\
12. Tomczyk, S. et al. Alfv\'en waves in the quiet solar corona. Science 317, 1192-1196 (2007)\\
13. Erdélyi, R., Taroyan, Y. Hinode EUV spectroscopic observations of coronal oscillations. Astron. 
Astrophys. 489, L49 (2008)\\
14. McIntosh, S. et al. Alfv\'enic waves with sufficient energy to power the quiet solar corona and fast 
solar wind. Nature 475, 477-480 (2011)\\
15. Braginskii, S. Transport Processes in a Plasma. Rev. Plasma Phys. 1, 205 (1965)\\
16. Morton, R., Erdélyi, R., Jess, D., Mathiudakis, M. Observations of sausage modes in magnetic pores. 
Astrophys. J. 729, L18 (2011)\\
17. Liu. W. et al. Direct Imaging of Quasi-periodic Fast Propagating Waves of ~ 2000 km s-1 in the Low 
Solar Corona by the Solar Dynamics Observatory Atmospheric Imaging Assembly. Astrophys. J. 736, L13 
(2011)\\
18. Shen, Y., Liu, Y. Observational study of the quasi-periodic fast propagating magnetosonic waves and 
the associated flare on 2011 May 30. Astrophys. J. 753, 53 (2012)\\
19. Jess, D. et al. The origin of type-I spicule oscillations. Astrophys. J. 744, L5 (2012)\\
20. Jess, D. et al. ROSA: A high-cadence, synchronized multi-camera system. Sol. Phys. 261, 363-373 
(2010)\\
21. Leenaarts, J. et al. DOT tomography of the solar atmosphere VI: Magnetic elements as bright points in 
the blue wing. Astron. Astrophys. 449, 1209-1218 (2006)\\
22. Leenaarts, J., Carlsson, M., van der Voort, L., The formation of the Hα line in the solar chromosphere, 
749, 136 (2012)\\
23. Heinzel, P., Schmieder, B. Chromospheric fine structure: black and white mottles. Astron. Astrophys. 
282, 939-954 (1994)\\
24. Spruit, H. Propagation speeds and acoustic damping of waves in magnetic flux tubes. Sol. Phys. 75, 
3-17 (1982)\\
25. Edwin, P., Roberts, B. Wave propagation in a magnetic cylinder. Sol. Phys. 88, 179-191 (1983)\\
26. Fedun, V., Sheylag, S., Erdélyi, R. Numerical modelling of footpoint-driven magneto-acoustic wave 
propagation in a localized solar flux tube. Astrophys. J. 727, 17 (2011)\\
27. Cally, P. Leaky and non-leaky oscillations in magnetic flux tubes, Sol. Phys. 103, 277-298 (1986)\\
28. Trujillo-Bueno, J. et al. The Hanle and Zeeman effects in solar spicules: a novel diagnostic window on 
chromospheric magnetism. Astrophys. J. 619, L191-L194 (2005).\\
29. Ionson, J. Resonant absorption of alfv\'enic surface wave and the heating of solar coronal loops. 
Astrophys. J. 226, 650-673 (1978)\\
30. Hollweg, J. Resonances of coronal loops. Astrophys. J. 277, 392-403 (1984)\\
31. Poedts, S., Goossens, M., Kerner, W. Numerical simulation of coronal heating by resonant absorption 
of Alfv\'en waves. Sol. Phys. 123, 83-115 (1989)\\
32. Ruderman, M., Berghmans, D., Goossens, M., Poedts, S. Direct excitation of resonant torsional 
Alfv\'en wave by footpoint motions. Astron. Astrophys. 320, 305-318 (1997)\\
33. Matthaeus, W. et al. Coronal heating by magnetohydrodynamic turbulence driven by reflected low-
frequency waves. Astrophys. J. 523, L93-L96 (1999)\\
34. Cranmer, S., van Ballegooijen, A., Edgar, R. Self-consistent coronal heating and solar wind 
acceleration from anisotropic magnetohydrodynamic turbulence. Astrophys. J. 171 (suppl), 520-551 
(2007)\\
35. Beliën, A., Martens, P., Keppens, R. Coronal heating by resonant absorption: the effects of 
chromospheric coupling. Astrophys. J. 526, 478-493 (1999)\\
36. Withbroe, G., Noyes, R. Mass and energy flow in the solar chromosphere and corona. Ann. Rev. 
Astron. Astro. 15, 363-387 (1977)\\
37. Aschwanden, M., Winebarger, A., Tsiklauri, D., Hardi, P. The coronal heating paradox. Astrophys. J. 
659, 1673-1681 (2007)\\
38.  Rimmele, T., Recent advances in solar adaptive optics. Society of photo-optical instrumentations 
engineers conference series. 5490, 34-46 (2004) \\
39. Wöger, A., von der Lühe, O., Reardon, K. Speckle interferometry with adaptive optics corrected solar 
data. Astron. Astrophys. 488, 375-381 (2008) \\
40. Jess, D. et al. High frequency oscillations in a solar active region observed with the Rapid Dual 
Imager. Astron. Astrophys. 473, 943-950 (2007)\\

\noindent \textbf{Supplementary Information} is linked to the online version of the paper at 
www.nature.com/nature

\noindent Supplementary Figs. S1 to S8\\
Supplementary Tables S1 to S3\\
Supplementary Discussion\\
Supplementary Methods\\
Supplementary References (41-43)\\
Supplementary Movies S1 to S4\\

\noindent\textbf{Acknowledgements} RE acknowledges M. Kéray for patient encouragement. The authors are also 
grateful to NSF, Hungary (OTKA, Ref. No. K83133), the Science and Technology Facilities Council (STFC), 
UK and the Leverhulme Trust for the support they received. Observations were obtained at the National 
Solar Observatory, operated by the Association of Universities for Research in Astronomy, Inc. (AURA), 
under agreement with the National Science Foundation. We would like to thank the technical staff at DST 
for their help and support during the observations and the Air Force Office of Scientific Research, Air 
Force Material Command, USAF for sponsorship under grant number FA8655-09-13085.

\noindent\textbf{Author Contributions} R.J.M (with D.K. and R.E.) performed analysis of observations. 
R.J.M, G.V., R.E. and M.S.R. interpreted the observations. D.B.J and M.M. performed all image processing. 
R.E. (Principal Investigator of the observations), M.M. and D.B.J. designed the observing run. All authors 
discussed the results and commented on the manuscript. 

\noindent\textbf{Author Information} Reprints and permissions information is available at 
www.nature.com/reprints. The authors declare no competing financial interest. Correspondence and 
requests for materials should be addressed to R.E. (robertus@sheffield.ac.uk).

\newpage

\centering{\large{\textbf{Supplementary material for}}}\\
\vspace{1cm} \centering{\large{\textbf{Observations of ubiquitous
compressive waves in the Sun's
chromosphere}}}\\
\vspace{0.5cm}\normalsize \centering{R. J. Morton, G. Verth, D. B.
Jess, D. Kuridze, M. S. Ruderman, M.
Mathioudakis, R. Erd\'elyi$^*$}\\
\vspace{0.1cm} \flushleft
$^{*}$ To whom correspondence should be addressed\\
\vspace{0.5cm}
\newpage

\Large \textbf{Supplementary Figures} \normalsize
\begin{figure*}[!htbp]\label{fig:fov}
\centering
\includegraphics[scale=0.95,clip=true, viewport=0.0cm 0.0cm 18.0cm 7.5cm]%
{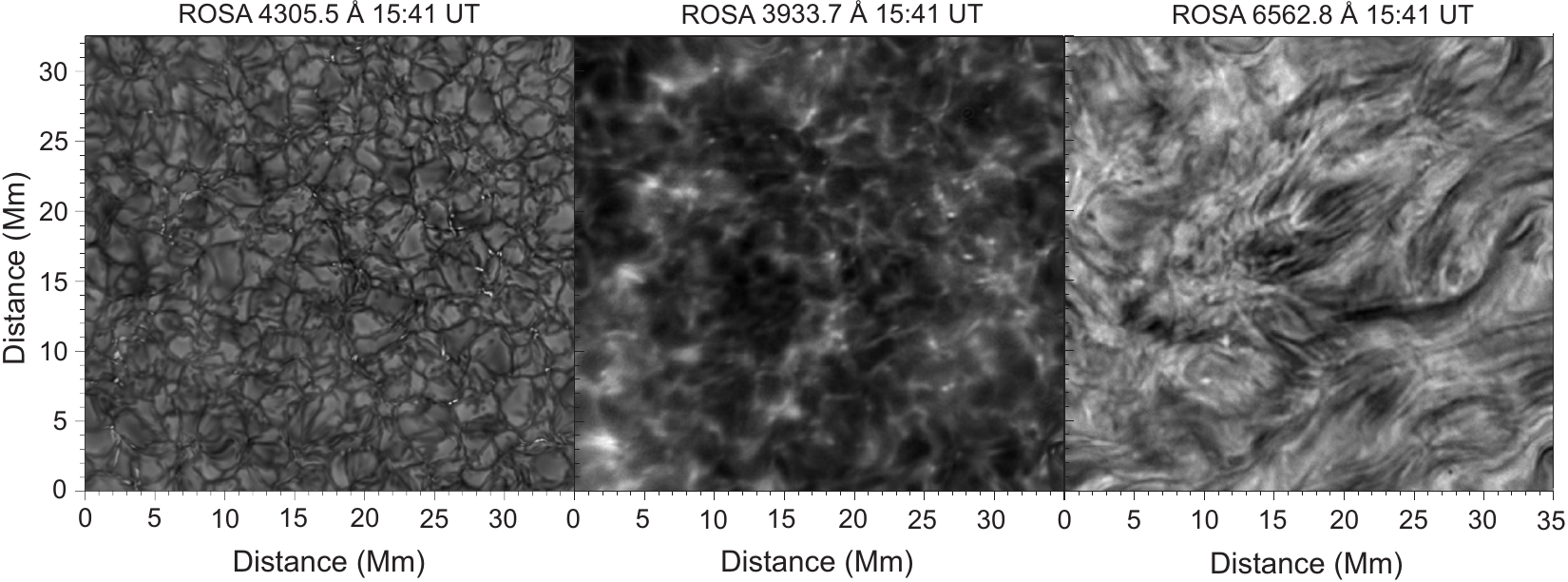} \caption{Supplementary Figure S1 \textbf{The photospheric magnetic
field:} The \textbf{left} panel shows a ROSA G-band image which
samples the lower photosphere. The bright points in the
intergranular lanes highlight strong concentrations of magnetic
flux. The \textbf{middle} panel is a co-temporal Ca II K image
showing the upper photosphere/low chromosphere. The magnetic bright
points seen in the G-band are also visible but have a more diffuse
appearance. The \textbf{right} panel shows the corresponding
H$_\alpha$ image. All images are co-temporal and co-spatial.}
\end{figure*}
\begin{figure*}[!htbp]
\centering
\includegraphics[scale=0.7]%
{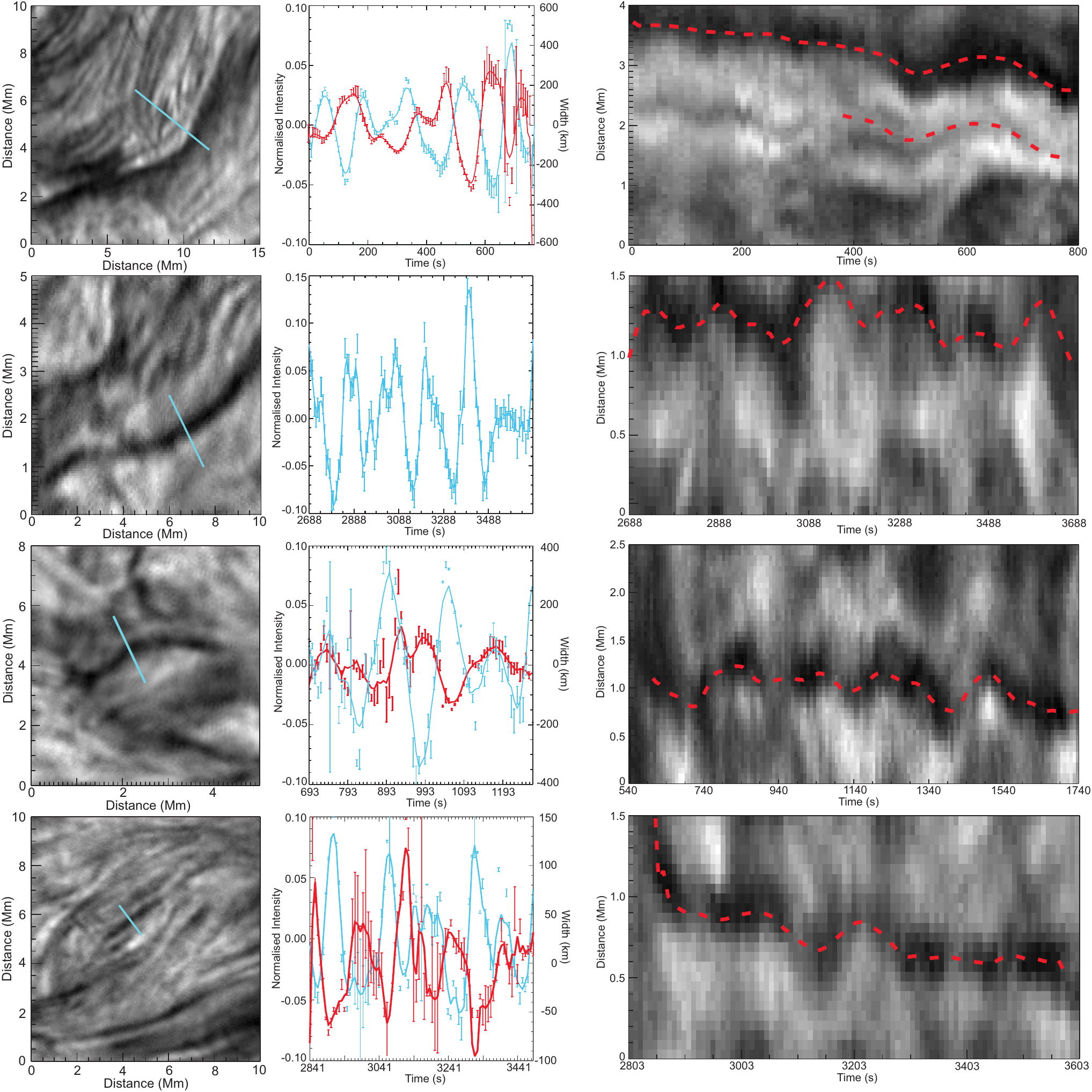}  \caption{Supplementary Figure S2 \textbf{Concurrent wave
modes:} \textbf{Left} panels show the H$\alpha$ chromospheric
magnetic structures in which the waves were observed. The images, in
descending order, are taken at $t=380$~s, $3055$~s, $770$~s,
$3095$~s from the beginning of the data series. \textbf{Middle}
panels display the structure's intensity (blue) and width (red) as a
function of time. The periodic behaviour in both intensity and width
demonstrates the compressive behaviour identified as the fast
sausage mode. \textbf{Right} panels show the observed fast kink
wave. The given distances start from the top of the cross-cuts shown
in the first panel. Details on each of the waves are given in Table
S1 and how the periodic phenomenon are measured is described in
Sections~2-4. Note, periodic phenomena shown in the second row have
not been derived from a Gaussian fit. To obtain the transverse
displacement of the structures axis we used the position of the
minimum value of intensity for each cross-sectional flux profile.
This technique has a larger error ($\pm50$~km) than the Gaussian fit
method. The value of minimum intensity is plotted in the second
panel.}
\end{figure*}
\newpage

\begin{figure*}[!htbp]
\centering
\includegraphics[scale=0.7,clip=true, viewport=0.0cm 0.0cm 10.0cm 8.5cm]%
{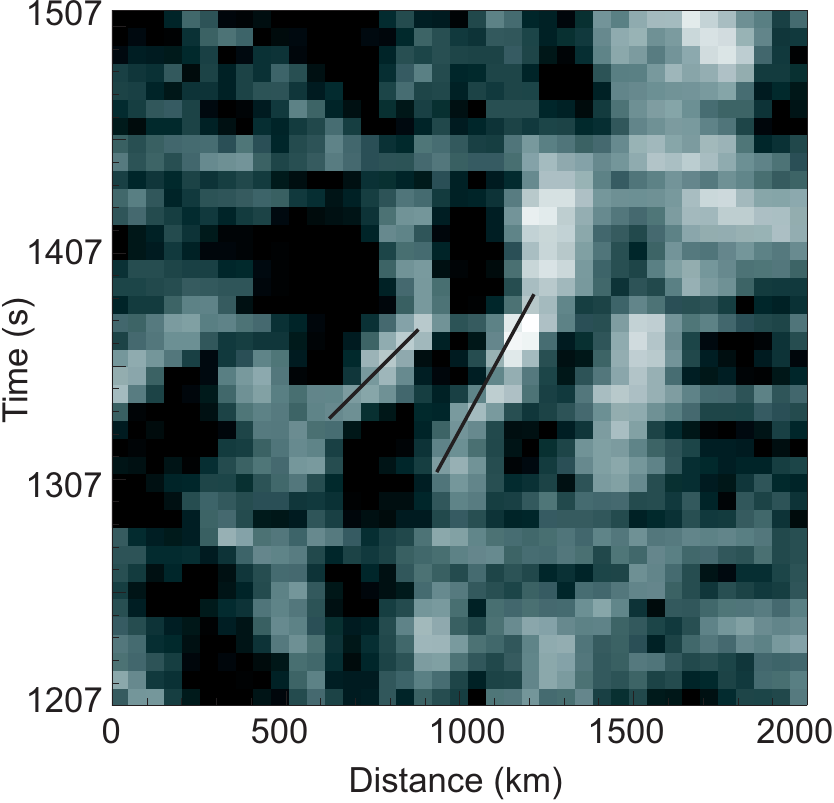}  \caption{Supplementary Figure S3 \textbf{Measuring transverse
motions:} A
time-distance diagram displaying transverse wave motion. The black
lines show typical linear fits to the transverse displacement.
Gradients of the lines are the time-averaged velocity amplitude.  }\label{fig:examp}
\end{figure*}

\begin{figure*}[!htbp]
\centering
\includegraphics[scale=0.7]%
{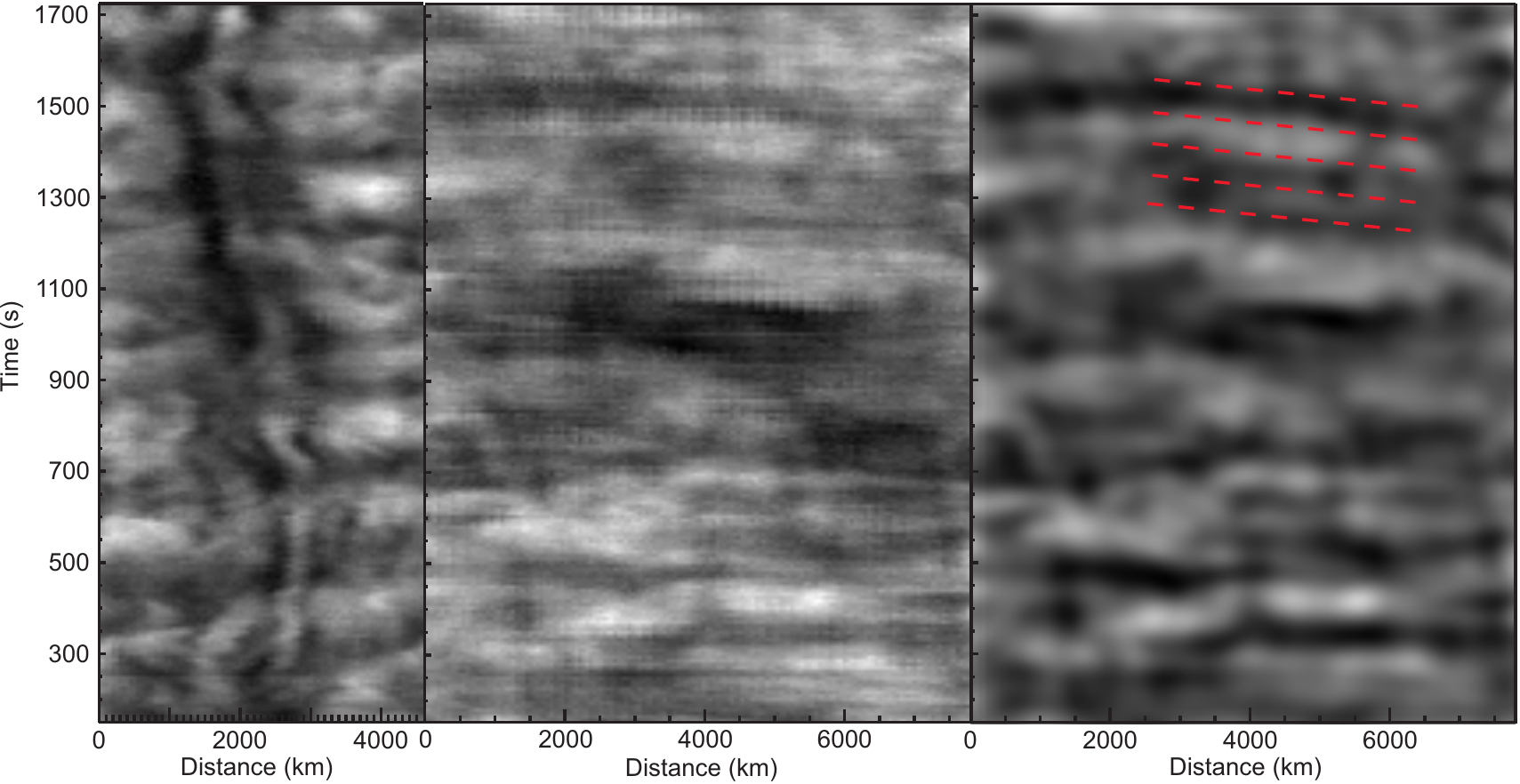}  \caption{Supplementary Figure S4 \textbf{Comparison of methods:}
The images shown here demonstrate the connection between the
compressive disturbances identified with Gaussian fits and the
periodic intensity phenomena seen in time-distance plots. The
\textbf{left} hand image shows the time-distance plot of the
chromospheric flux tube, discussed in Fig.~4, which demonstrated
concurrent wave motions using the Gaussian fitting. The
\textbf{middle} panel is a time-distance diagram for a cross-cut
parallel to the flux-tube (see, Supplementary Fig.~S5). The \textbf{right} panel displays the
high pass filtered images of the middle panel, showing the periodic,
propagating intensity disturbances (highlighted by the red dashed
lines) which were previously identified with the Gaussian fitting.
The measured propagation speed is $64\pm4$~km\,s$^{-1}$.
}\label{fig:fs6}
\end{figure*}

\begin{figure*}[!htbp]
\centering
\includegraphics[scale=0.75,clip=true, viewport=0.0cm 0.0cm 22.0cm 9.0cm]%
{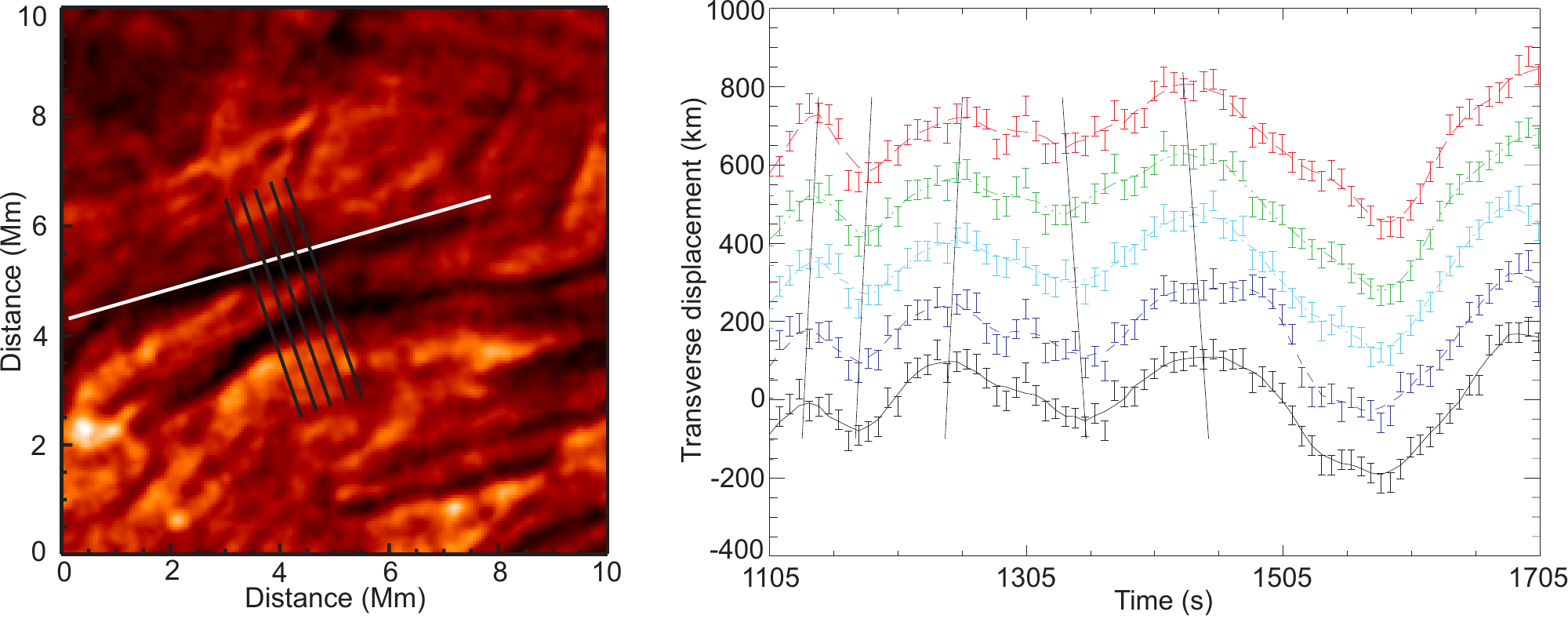}  \caption{ Supplementary Figure S5 \textbf{Measuring phase speeds:} The
\textbf{left} panel shows the same chromospheric structure as that shown in Fig.~4. The white line 
shows the position of the parallel cross-cut used in the middle panel of Supplementary Fig.~7. The 
\textbf{right} panel shows the transverse motions from cross-cuts (shown in the left panel) at
different positions along the chromospheric structure. The data points are obtained from the Gaussian 
fits and the error-bars are the one standard deviation error on the fit. The
diagonal lines highlight the $87$~km\,s$^{-1}$ upwards (first set of
lines) and $71$~km\,s$^{-1}$ downwards (second set) propagating
waves. The visible shift in the signals confirms the results
obtained from cross-correlation.} \label{fig:phase_n}
\end{figure*}

\begin{figure*}[!htbp]
\centering
\includegraphics[scale=0.9]{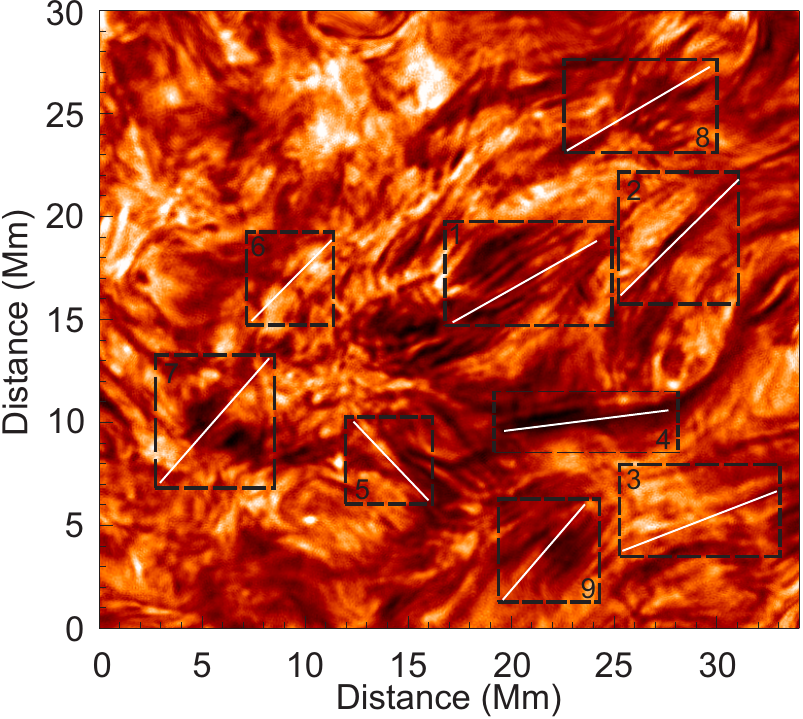}  
\caption{Supplementary Figure S6 \textbf{Regions showing fast
propagating, periodic intensity perturbations:} A sub-region of the
ROSA field of view. The boxes indicate areas in which evidence for
fast propagating disturbances are found and the white lines show the cross-cut positions. The details of 
the observed
oscillations are given in Supplementary Table S3.}\label{fig:fs7}
\end{figure*}

\begin{figure*}[!htbp]
\centering
\includegraphics[scale=0.9,clip=true, viewport=0.0cm 0.0cm 18.0cm 7.5cm]%
{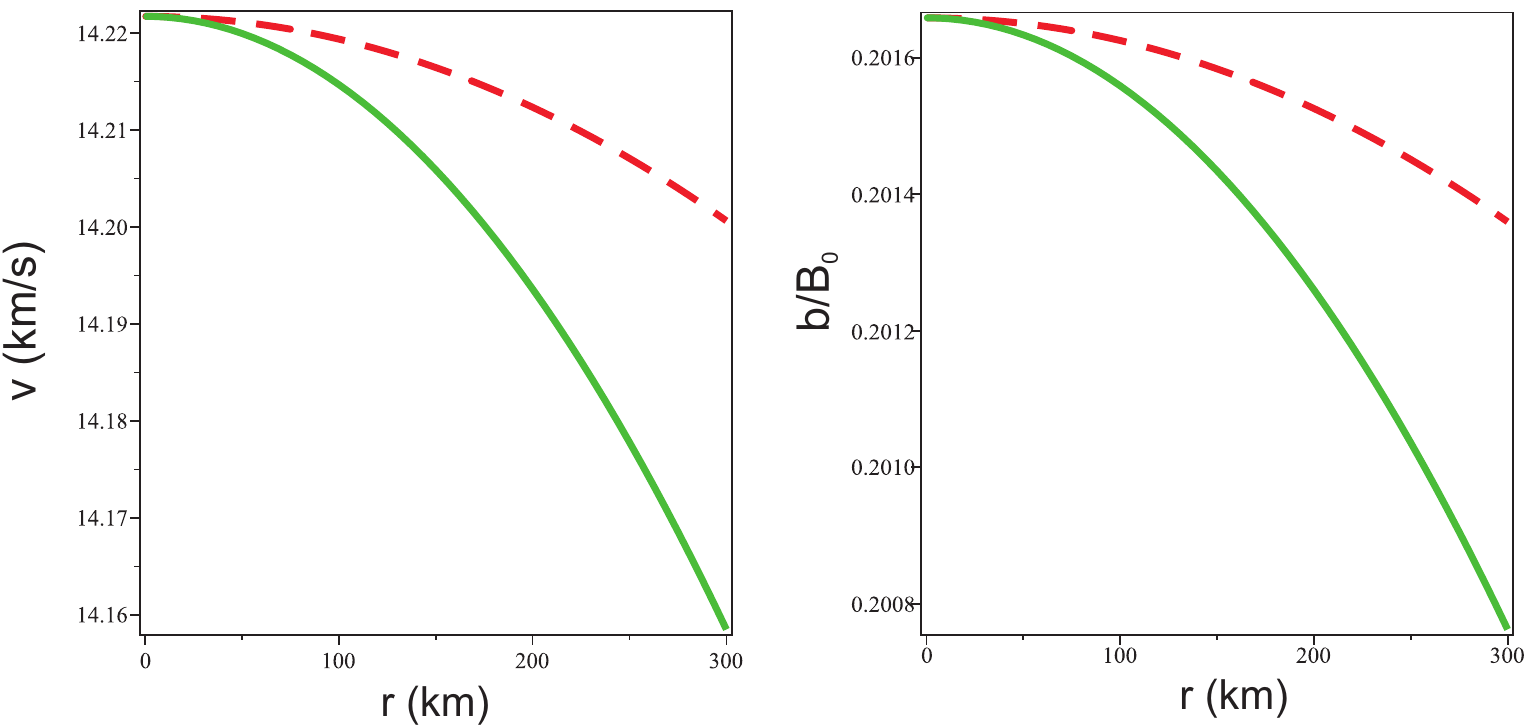}  \caption{Supplementary Figure S7 {\textbf{Theoretical amplitudes for
the fast kink mode:}} The \textbf{left} panel shows the amplitude of
the velocity perturbations as a function of radius: $v_r$ is the
solid and $v_{\theta}$ is the dashed line. The amplitude of the
magnetic field perturbations as a function of radius is demonstrated
in the \textbf{right} panel. The values of the perturbations are
normalised with respect to the equilibrium magnetic field, both of
which cannot be measured directly: $b_r$ is the solid and
$b_{\theta}$ is the dashed line. The amplitude distributions are
calculated for $c_{ph}=70$~km\,s$^{-1}$ and
$k=5\times10^{-4}$~km.}\label{fig:tkw}
\end{figure*}

\begin{figure*}[!htbp]
\centering
\includegraphics[scale=0.75,clip=true, viewport=0.0cm 0.0cm 18.0cm 9.0cm]%
{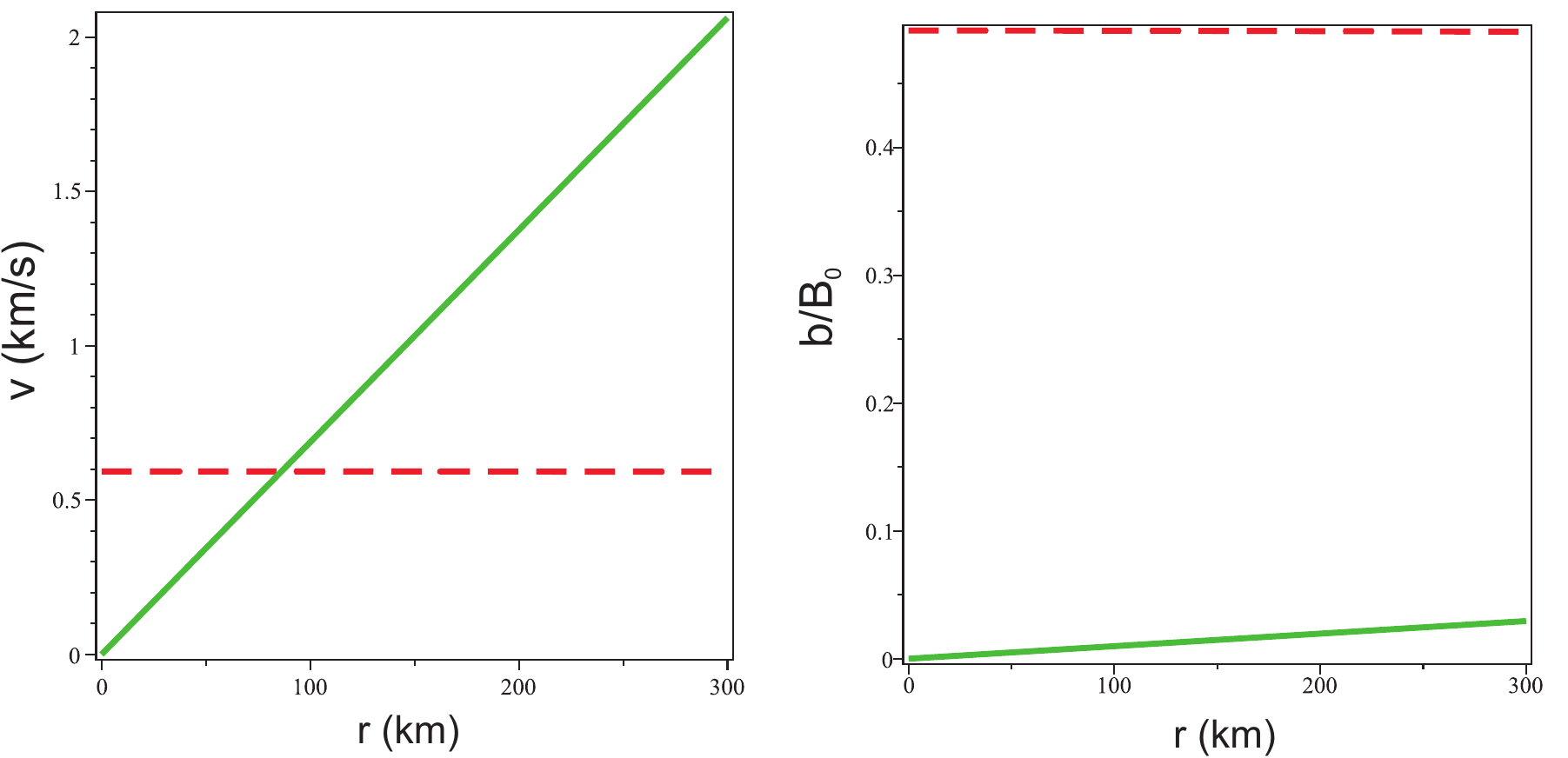}  \caption{Supplementary Figure S8 \textbf{Theoretical amplitudes
for the fast MHD sausage mode:} In the \textbf{left} panel $v_r$ is
the solid and $v_{z}$ is the dashed line. In the \textbf{right}
panel $b_r$ is the solid and $b_{z}$ is the dashed line. The values
of the perturbations are normalised with respect to the equilibrium
magnetic field, both of which cannot be measured directly. The
amplitude distributions are calculated for $c_{ph}=70$~km\,s$^{-1}$
and $k=4\times10^{-4}$~km. Note the different wave numbers of Figs.
4 and 5 that arises from the difference in observed periods and
phase speeds of the two MHD wave modes.} \label{fig:fsm}
\end{figure*}
\newpage

\vspace{0.5cm} \Large{\textbf{Supplementary Tables}} \normalsize
\begin{center}
\begin{tabular} {|l | c | c | c | r|}
\hline &\multicolumn{2}{|c|}{Fast kink} & \multicolumn{2}{|c|}{Fast
MHD Sausage}
\\ \hline
&$c_{ph}$ (km $s^{-1})$ & Period (s) & $c_{ph}$ (km s$^{-1}$) & Period (s) \\
\hline 
1 & $80 \pm 20$ & $180\pm 5$ & $102\pm 15$   & $135 \pm 3$\\\hline 
   & $58 \pm 15$ &  &  & \\ \hline 
2 & $92 \pm 26$ & $180\pm 5$ & $110\pm 27$   & $214 \pm 6$\\ \hline 
3 & $58 \pm 17$ & $210\pm 3$ & $87\pm 26$     & $154 \pm 5$\\ \hline 
4 & -                    & $182\pm 5$ & $51\pm 12$     & $203 \pm 5$\\ \hline
\end{tabular}\label{tab2}

\end{center}
\textbf{Supplementary Table S1.} The measured properties of the observed waves in
Supplementary Fig.~S2. The numbers in the left column refer to the
row in Supplementary Fig.~S2. The wave labelled 1 displays both
upward and downward propagating kink waves so two phase speeds are
given.

\begin{center}
\begin{tabular} {|l | c |  r|}
\hline
Region & $c_{ph}$ (km s$^{-1})$ & Period (s) \\
\hline 1 &$135\pm17$ & $136\pm10$ \\
             &$270\pm135$& $130\pm10$ \\
             &$78\pm9$& $90\pm10$ \\
\hline 2 &$86\pm8$& $150\pm10$ \\
             &$127\pm12$& $150\pm10$ \\
\hline 3 &$152 \pm 15$ & $140\pm10$  \\
             &$122\pm25$& $146\pm10$ \\
             &$257\pm61$& $148\pm10$ \\
\hline 4 &$246 \pm 80$ & $121\pm10$ \\
             &$82\pm14$& $136\pm10$ \\

\hline 5& $48 \pm 8$ & $100\pm 10$ \\
          &$325\pm81$& $100\pm10$ \\

\hline 6& $55\pm5$& $116\pm10$ \\

\hline 7 &$49\pm4$& $162\pm10$ \\
             &$84\pm21$& $108\pm10$ \\

\hline 8 &$57\pm3$& $94\pm10$ \\
\hline 9 &$58\pm10$& $130\pm10$ \\
             &$55\pm4$& $192\pm10$ \\
       
          \hline
\end{tabular}\label{tab2}

\end{center}
\textbf{Supplementary Table S2.} The measured properties of the fast propagating,
periodic intensity perturbations. The regions are shown in
Supplementary Fig.~S6.

\begin{center}
\begin{tabular} {|l | c|}
\hline
T & 10,000~{K}\\
$B_i$ & 0.001-0.002~{Tesla}\\
$\rho_i$ & 2-3$\times10^{-10}$~kg\,m$^{-3}$\\
$v_A=B/\sqrt{\mu_0\rho}$ & $50-100$~km\,s$^{-1}$\\
$c_s=\sqrt{R T/\tilde{\mu}}$ & $10$~km\,s$^{-1}$\\
\hline
\end{tabular}\label{tab1}

\end{center}
\textbf{Supplementary Table S3.} The typical chromospheric plasma parameters used
for calculations of the wave energy. Here $T$ is the plasma
temperature (ref. \textit{1,22,23}), $B$ is the magnetic field (ref.
\textit{28}), $\rho$ is the chromospheric density (refs.
\textit{1,22,23}), $v_A$ is the Alfv\'en speed, $\mu_0$ is the
magnetic permeability of free space, $c_s$ is the sound speed, $R$
is the gas constant and $\tilde{\mu}$ is the mean atomic weight. The
subscript $i$ refers to the internal quantities of the chromospheric
structure.

\newpage
\vspace{0.5cm} \Large{\textbf{Supplementary Discussion}} \normalsize

\vspace{0.2cm} \noindent\textbf{Magnetic field structure}

We provide additional ROSA data here in order to put our
chromospheric observations in the context of the underlying magnetic
field. The ROSA G-band (Supplementary Fig.~S1 left panel) filter
reveals a number of bright features in the inter-granular lanes.
These features are frequently referred to as magnetic bright points
(MBPs). The MBPs are common throughout the quiet Sun and usually
correspond to strong magnetic field concentrations ($\sim1$~kGauss)
(ref. \textit{41}). The MBPs highlight the super-granule network and
are thought to correspond to open magnetic fields that reach into
the corona (e.g., ref. \textit{42}). The Ca II K line samples the
lower chromosphere but contains significant contributions from
photospheric sources. The MBPs are also seen in the Ca II K images
(Supplementary Fig.~S1 middle panel), however their emission appears
more diffuse. Previous observations (e.g., ref. \textit{42}) have
demonstrated that fibrils and mottles (seen in  H$\alpha$) extend
out from the strong flux concentrations.

\vspace{0.2cm} Typically, flux tubes undergo expansion between the
photosphere and chromosphere, hence this will reduce the magnitude
of the magnetic field at chromospheric level compared to the
photospheric value. Spectropolarimetric observations (ref.
\textit{28}) have shown that the typical field strength from the
chromosphere at the limb is of the order $10-40$~G
($0.001-0.004$~Tesla). Using this value of magnetic field and
typical values of density measured for chromospheric magnetic
structures, we obtain estimates of Alfv\'en speeds (Supplementary
Table S1) that are comparable to the phase speeds of the waves
observed here.

\vspace{0.5cm} We can set an upper limit on the area of the
chromosphere occupied by open magnetic fields. If we assume each of
the 300 chromospheric structures identified (Fig.~1) is formed along
an open field line, i.e., protrudes into the corona, then the
cross-sectional area of the observed structures would only cover
approximately $4-5\%$ of the observed chromosphere. This percentage
is in agreement with estimates from previous observations (refs.
\textit{1, 36} ). Of course, not all the observed structures are
open but this limit demonstrates only a small percentage of the
chromosphere is open to the corona. This has implications for the
wave-energy able to reach the corona from the chromosphere as
discussed in the main paper.

We note that this estimate of connectivity between the chromosphere
and corona is constrained by the current spatial resolution and the
ability to resolve very fine structure. However, we suggest this
upper bound would not change much even with improved resolution. At
present, chromospheric structures with widths on the order of
$100$~km are reported. These occupy less than a tenth of the volume
of structures with the median width of $360$~km. Hence, a large
number of unresolved structures need to exist to significantly alter
the upper bound placed on the connection between the chromosphere
and corona.

\vspace{0.2cm} \noindent\textbf{Explaining the compressive perturbations}

The determination of which wave mode causes the observed periodic
changes in intensity and cross-section is challenging. We begin by
ruling out some of the possibilities.
\begin{itemize}
\item[$\bullet$]The observed modes cannot be slow modes as they propagate with speeds
close to the local Alfv\'en speed. Slow modes propagate with phase
speed close to the sound speed (ref. \textit{25}).
\item[$\bullet$]An alternative option is that we detect a kink
mode polarised in the line-of-sight in addition to the transverse
displacement in the observational plane. This would, however, invoke
a helical motion which has previously been tentatively reported in
spicules. During a helical motion, both polarisations would have the
same phase speed and period.

We use a narrowband H$\alpha$ filter, so Doppler shifts caused by
the line-of-sight motion could cause an increase in intensity.
However, the intensity would increase whether the flux tube was
moving towards or away from the observer. Hence, if we had two
oppositely polarised kink modes, the period for the change in
intensity due to  line-of-sight motion should be half the value of
that measured for the observed transverse displacement. This is not
the case. Secondly, this motion would not explain the change in the
visible width of the tube. The entire flux tube would move towards
or away from the observer at the same speed, leaving the visible
width unaffected.
\item[$\bullet$]The linear torsional Alfv\'en wave can also be ruled out as it is
incompressible and does not cause a change in the flux tube width.
\item[$\bullet$] The next possibility is that the observed change in intensity and width is
a non-linear wave phenomenon.
\begin{itemize}
\item[$\bullet$]Recent work (ref. \textit{43}) on propagating non-linear kink
motions has shown that the non-linearity couples the kink motion to
higher order (fluting) modes which can distort the cross-section of
the flux tube. These higher order modes propagate with the same
phase speed as the kink mode. Non-linear theory suggests that the
higher order modes are relatively incompressible, which would leave
the intensity perturbations unexplained.

\item[$\bullet$]The non-linear torsional Alfv\'en wave causes a compression of the
cross-section and may alter the pressure inside the loop. This
compression would have half the period of the non-linear Alfv\'en
wave since compression would occur as the motion twists the magnetic
surface in either direction. This implies the period of the
non-linear Alfv\'en motion would be $\sim400$~s. However, we can
discard the torsional Alfv\'en wave as it would propagate with the
{\it internal} Alfv\'en speed that is substantially smaller than the
kink speed. In fact, our observations demonstrate just the opposite
relation.
\end{itemize}
\end{itemize}
The crucial point from the observations is that the perturbations in width
and intensity show a distinct phase relation. This suggests that the perturbations
are related and can be attributed to the same wave mode. Since we have already ruled
the role Doppler effects play in the observed behaviour, we conclude here that the
perturbations are genuinely a compressive phenomenon.

The only remaining choice is the fast sausage MHD mode. However, a
trapped mode cannot propagate unless the dimensionless value $ka$
satisfies (refs. \textit{25, 27})
$$
ka>j_0\left(\frac{v_A^2}{v_{Ae}^2-v_A^2}\right)^{1/2},
$$
where the relationship is derived for low-beta plasma, and
$j_0\approx2.4$ is the first zero of the Bessel function $J_0$. The
definition of a trapped mode is a wave that decays exponentially away
from the wave guide.

To obtain the lowest possible cut-off value, we assume a cold, dense
chromospheric structure embedded in a coronal plasma thus providing
a maximum possible density contrast. Hence, $B_i\approx B_e$ and
$\rho_i\gg\rho_e$, where $\rho_i=3\times10^{-10}$~kg\,m$^{-3}$ (a
typical value for chromospheric structures) and
$\rho_e\sim10^{-12}$~kg\,m$^{-3}$ (a typical coronal value), then
the dimensionless wavenumber has to satisfy $ka>0.2$. The wavenumber
for the observed oscillation in Fig. 4 is
$k\approx3.8\pm1.0\times10^{-4}$~km$^{-1}$ and $a=200\pm50$~km,
giving a predicted value of $ka\sim0.08\pm0.03$. Even with these
extreme density contrast values the observed $ka$ is still less than
the cut-off value. Therefore, we are confident that we observed a
leaky fast sausage MHD wave.

Leaky modes differ from trapped modes in that their wave energy is
radiated away from the tube as the wave propagates, typically
resulting in short lifetimes on the order of a period. Because we
see multiple periods of a propagating mode this may suggest a
continuous driver. Since the wave energy leaks from the tube, the
surrounding chromospheric structures should experience coherent
oscillatory behaviour, which is seen in Supplementary Movies S3 and S4, where a
number of chromospheric magnetic structures appear to support fast
propagating disturbances at approximately the same time.

\vspace{0.5cm} \Large{\textbf{Supplementary Methods}} \normalsize

\vspace{0.2cm} \noindent\textbf{Measuring the period of the waves}

The period of the observed oscillatory phenomena is measured using wavelet analysis. The observed 
oscillation in either central position/intensity/width is subject to a wavelet transform and 
the wavelet component with the greatest power is identified as the period of the oscillation. The wavelet 
decomposes the signal at certain frequency intervals. The size of the wavelet scales used means there is 
a 20~s uncertainty in the determining the period. The error is reduced by making multiple measurements 
of the period from different cross-cuts from the Gaussian fits (see, e.g. Supplementary Table~1). For the 
slits placed parallel to the fine structure (see, e.g. Supplementary Table~2), we measure the period of the 
oscillation in intensity in five columns in each cross-cuts through the oscillations and take the average 
value. The error on the average period is then the reduced value of $\sim10$~s.

\vspace{0.2cm} \noindent\textbf{Calculating the wave energy}

Here we demonstrate how the wave energy flux can be calculated for a
waveguide model of the chromosphere. The waveguide model attempts to
take into account the observed fine-structure of the chromosphere.

The energy flux of the observed waves can be estimated from the
measured amplitudes and phase speeds. The first assumption is that
the chromospheric magnetic structures are over-dense homogeneous
cylindrical magnetic waveguides embedded in an ambient plasma. To
reduce the complexity of the calculations, here we use linear theory
to relate the amplitudes of perturbations of different quantities.
The waveguide supports an infinite set of orthogonal MHD wave
eigenmodes. In the non-linear theory (ref. 43), eigenmodes strongly
interact and cannot be separated. Linear theory provides a good
first approximation of the wave energy flux.

We are interested in the energy flux at a particular height in the
solar atmosphere, so we neglect the variation of background
quantities along the waveguide and assume that the plasma is
homogeneous in the radial direction inside. Using cylindrical
coordinates $(r,\theta,z)$ and assuming that the perturbations of
all quantities are proportional to $\exp{-i(\omega t-n\theta-kz)}$,
we obtain the following set of equations describing the dependence
of perturbations on the radial coordinate $r$ (see, e.g., ref.
\textit{24})
\begin{eqnarray}\label{eq:amps}
v_r&=&-A\frac{\omega^2-k^2c_s^2}{m_0^2\omega^2}\frac{dJ_n(m_0r)}{dr},\hspace{1.5cm}
v_{\theta}=Ai\frac{\omega^2-k^2c_s^2}{m_0^2\omega^2}\frac{n}{r}J_n(m_0r),\nonumber\\
v_z&=&-Ai\frac{kc_s^2}{\omega^2}J_n(m_0r),
\hspace{2.75cm}b_r=-\frac{k}{\omega}B_0v_r,\hspace{3.0cm}\mbox{(S1)}\nonumber\\
b_\theta&=&\frac{k}{\omega}B_0v_\theta,\hspace{4.2cm}
b_z=Ai\frac{\omega^2-k^2c_s^2}{\omega^3}B_0J_n(m_0r).\nonumber
\end{eqnarray}
Here, ${\bf v}=(v_r,v_{\theta},v_z)$ are the velocity perturbations,
${\bf b}=(b_r,b_{\theta},b_z)$ are the magnetic field perturbations,
$\omega$ is the frequency of the oscillation, $k$ is the wavenumber,
$J_n$ is the Bessel function of the first kind, $n$ refers to the
angular wave number (or mode number) and $A$ is some constant which
is used to scale the amplitudes. Further,
$$
m_0^2=\frac{(\omega^2-v_A^2k^2)(\omega^2-c_s^2k^2)}{(cs^2+v_A^2)(\omega^2-c_T^2k^2)},
$$
where $c_T=c_sv_A/\sqrt{(c_s^2+v_A^2)}$ is the tube speed. The use
of $J_n$ is only valid as long as  $c_T<c_{ph}<\min{(c_s,v_A)}$ or
$\max{(c_s,v_A)}<c_{ph}$, which guaranties that $m_0^2>0$. In this
case the wave is referred to a body mode. If these conditions are
not satisfied, and $m_0^2<0$, then we have to substitute
$I_n(|m_0|r)$ for $J_n(m_0r)$. In this case the wave is referred to
as a surface wave.

\vspace{0.2cm}\textbf{Fast kink wave}\\
Using the set of Equations~(S1) with $n=1$ and the
estimates for the plasma parameters, we determine the amplitudes of
the perturbations as a function of radius.

In Supplementary Fig.~S7 we show the amplitude profiles of $v_r$,
$v_\theta$, $b_r$ and $b_\theta$ as a function of radius. The
profiles are shown for a waveguide of radius $300$~km, which is
typical for chromospheric structures (see, Fig.~1). For the plots in Supplementary Fig.~S7 a 
phase speed $c_{ph}=70$~km\,s$^{-1}$ is used and the
wavenumber $k=5\times10^{-4}$~km is calculated using the phase speed
and the typical periods of the kink oscillations, $P=180-200$~s. The
amplitudes vary only by $<1\%$ across the internal plasma of the
waveguide.

Next, we assume that the observed waves are propagating in a cold
plasma, i.e., the plasma-$\beta$ is small, where $\beta$ is the
ratio of the gas pressure to the magnetic pressure. For the given
values of plasma parameters $\beta<0.03$ (see, Supplementary
Table~S3), so this approximation is valid. The equation for the time
averaged energy flux is then given by
\begin{equation}
<E>=\frac{1}{4}c_{ph}\left[\rho_i(v_r^2+v_\theta^2+v_z^2)
+\frac{1}{\mu_0}(b_r^2+b_\theta^2+b_z^2)\right].\hspace{3.0cm}\mbox{(S2)}\nonumber
\end{equation}
The first set of terms in the squared brackets is the kinetic energy
flux, while the second set relates to the magnetic energy flux. This
equation requires that we know the amplitudes of the perturbations.
However, we have measured the time-averaged value of the velocity
for the transverse motions. Hence, to calculate the time-averaged
energy flux using the measured velocity we use
\begin{eqnarray}
<E>&=&\frac{1}{2}c_{ph}\left[\rho_i(<v_r>^2+<v_\theta>^2+<v_z>^2)
+\frac{1}{\mu_0}(<b_r>^2+<b_\theta>^2+<b_z>^2)\right]. \nonumber \\
& &\hspace{11.5cm}\mbox{(S3)}\nonumber
\end{eqnarray}

For the fast kink wave the $v_z$ and $b_z$ perturbations are
sufficiently small and may be neglected. By substituting the
calculated average values of the perturbations into the energy flux
equation, we obtain a range of the energy flux for the transverse
mode, $<E_{k}>=4300\pm2700$~Wm$^2$. Here we have used
$<v_{r}>=5-10$~km\,s$^{-1}$ and the range of values given in
Supplementary Table~S3. We note the energy of the modes is evenly distributed
between the magnetic and kinetic terms.

\vspace{0.2cm} \noindent\textbf{Fast MHD sausage mode}

To obtain the perturbation amplitudes for the fast MHD sausage
modes, we set $n=0$ in the set of Equations~(S1). This
decouples the $v_\theta$ and $b_{\theta}$ perturbations from the
other perturbations, hence, the sausage mode has no azimuthal
components. To determine the amplitude profiles we set the amplitude
at the tube boundary equal to the value estimated from the
observations. Note, the $v_r$ estimated for the fast MHD sausage
mode would be the amplitude and not the time-averaged value.

The calculated amplitude profiles for the fast MHD sausage mode are
shown in Supplementary Fig.~S6. The $v_r$ and $b_r$ perturbations
are zero at the centre of the waveguide and reach a maximum at its
boundary. On the other hand, the amplitudes of the $v_z$ and $b_z$
perturbations are approximately constant across the waveguide
radius. It also worth noting that the amplitudes of the $b_z$
perturbations for the fast MHD sausage mode are large compared to
the $b_r$ components.

The equation for the time-averaged energy flux for the fast MHD
sausage mode in a low-$\beta$ plasma is
\begin{equation}
<E_{s}>=\frac{1}{4}c_{ph}\left[(v_r^2+v_z^2)+(b_r^2+b_z^2)\right].
\hspace{3.0cm}\mbox{(S4)}\nonumber
\end{equation}
The estimated amplitudes for $v_r$ are small, e.g. when compared to
the velocity amplitudes of the fast kink mode, and the dominant
terms are due to the magnetic energy flux. The estimated wave energy
flux for the fast sausage mode is in the range
$E_{s}=11700\pm3800$~Wm$^{-2}$ for velocity amplitudes of
$1.5-2$~km\,s$^{-1}$ and the range of values given in Supplementary
Table~S3.

\normalsize \vspace{1.0cm} \Large \textbf{Supplementary References}
\normalsize \noindent

\noindent 41. Berger, T., Title, A., On the relation of G-band
bright points to the photospheric magnetic field. \textit{Astrophys.
J.} \textbf{533}, 449-469 (2001)

 \noindent42. Rutten, R. J., Observing the solar
chromosphere. \textit{The physics of chromospheric plasmas ASP
conference series, Ed. P. Heinzel, I. Dorotovi\v{c}, R. J. Rutten.}
\textbf{368}, 27 (2007)

\noindent43. Ruderman, M. S., Goossens, M., Andries, J. Nonlinear
propagating kink waves in thin magnetic tubes. \textit{Phys.
Plasmas} \textbf{17}, (2010)

\end{document}